\documentclass[12pt]{article}
\renewcommand{\abstractname}{\ }
\begin{document}
\renewcommand{\abstractname}{\ }

\title{A refinement of the Schr\"odinger equation involving dissipation of waves in the phase space}
\author{E. M. Beniaminov}
\date{}
\maketitle
\begin{abstract}
{We give an example of a mathematical model describing quantum mechanical processes interacting with medium.
 As a model, we consider the process of heat scattering of a wave function defined on the phase space.
We consider the case when the heat diffusion takes place only with respect to momenta.
We state and study the corresponding modified Kramers equation for this process.
We consider the consequent approximations to this equation in powers of the quantity inverse to
the medium resistance per unit of mass of the particle in the process. The approximations are constructed
similarly to statistical physics, where from the usual Kramers equation for
the evolution of probability density of the Brownian motion of a particle in the phase space, one
deduces an approximate description of this process by the Fok\-ker--Planck
equation for the density of probability distribution in the configuration space.

 We prove that the zero (invertible) approximation to our model with respect to the large
 parameter of the medium resistance, yields the usual quantum mechanical description by the
 Schr\"odinger equation with the standard Hamilton operator.
We deduce the next approximation to the model with respect to the negative power of the medium resistance
coefficient. As a result we obtain the modified Schr\"odinger equation taking into account
dissipation of the process in the initial model.
}
\end{abstract}

\section{Introduction}
In this paper we continue the study of the generalized Kramers equation introduced in the paper \cite{ben_kramers}.

In \cite{ben_kramers}, the generalized Kramers equation arose as a mathematical model of the scattering process
of waves in the phase space under the action of medium in the heat equilibrium.
In \cite{ben_kramers} it has been shown that, for certain parameters of the model, the process described by the
generalized Kramers equation, is the composition of a rapid transitional process and a slow process.
The slow process, as proved in loc.~cit., is approximately described by the Schr\"odinger equation
used for description of quantum processes. Thus it has been shown that the quantum mechanical description
arises as the asymptotic description of the heat scattering process of waves in the phase space.

The purpose of the present paper is to construct an approximate equation describing the slow part
of the process given by the generalized Kramers equation, with the precision showing dissipative effects
in this process.

The paper consists of four Sections and the Appendix.

In the next Section we present the setting of the problem, the generalized Kramers equation for the heat scattering
process of waves on the phase space, and the properties of the operators involved in the generalized
Kramers equation.

In Section~3 we recall our earlier results describing the process given by the generalized Kramers equation.
We also give a method for the approximate description of this process in negative powers of the
medium resistance coefficient, and state the main result of the paper, The\-o\-rem~4.
In this Theorem we write down an equation describing the slow component of the studied process,
taking into account dissipation.
This equation is the modified Schr\"odinger equation, in which the process is not invertible.
We show that consequences of the modified Schr\"odinger equation are the effects of decoherence and
spontaneous jumps between the levels.

In Section~4 we state some further possible directions of research and possibilities of comparison of the
presented model with experiment.

Proof of the main Theorem 4 of the paper is given in the Appendix.

\section{Mathematical setting of the problem, and the main properties of the operators in the equation}

Thus, we consider the following mathematical model of a process. A state of the process at each moment of time
$t\in R$ is given by a complex valued function
$\varphi [x, p,t]$   on the phase space $(x,p) \in R^{2n}$,
where $n$ is the dimension of the configuration space.  Coordinates in the configuration space are
given by a tuple of numbers $x\in R^{n}$, and momenta by a tuple of numbers $p\in R^{n}$.
(Below everywhere we will write the arguments of a function in the square brackets,
in order to distinguish an argument of a function from multiplication by a variable.)

The generalized Kramers equation for a function $\varphi [x, p,t]$ is defined as the following equation:
\begin{equation}\label{eq_diff}
\frac{\partial\varphi}{\partial{t}}=A\varphi
+\gamma B{\varphi},
\end{equation}
\begin{equation}\label{def_A}
\mbox{where }\ \ \ \ \ \ A\varphi =\sum_{j=1}^{n}
\biggl(
\frac{\partial V}{\partial x_j} \frac{\partial\varphi}{\partial p_j}-
 \frac {p_j}{m} \frac{\partial \varphi}{\partial x_j}
\biggr)
-\frac{i}{\hbar}
\biggl(V-\sum_{j=1}^{n}\frac{p^2_j}{2m} \biggr)\varphi \ \ \ \ \ \ \
\end{equation}
\begin{equation}\label{def_B}
\mbox{and }\ \ \ \ \  B{\varphi}=
\sum_{j=1}^{n}\frac{\partial}{\partial{p_j}}\left( \biggl( p_j +i\hbar \frac{\partial}{\partial{x_j}}\biggr){\varphi}
+k_BTm\frac{\partial{\varphi} }{\partial{p_j}}\right); \ \ \ \ \ \ \ \nonumber
\end{equation}
$m$ is the mass of the particle;
$V[x]$ is the potential function of external forces acting on the particle;
$i$ is the imaginary unit;
$\hbar$ is the Planck constant;
$\gamma=\beta /m$ is the medium resistance coefficient $\beta$ per unit of mass of the particle;
$k_B$ is the Boltzmann constant; $T$ is the temperature of the medium.

If we pass in equation (\ref{eq_diff}) to the following dimensionless variables:
\begin{equation}\label{new_var}
 p'=\frac{p}{ \sqrt {k_BTm}}, \ \ \ \ x'=\frac{ \sqrt {k_BTm}}{\hbar}x,\ \ \ \ V'[x]=\frac{V[x]}{k_BT},
\end{equation}
then in the new variables equation (\ref{eq_diff}) takes the following form:
\begin{equation}\label{eq_diff'}
\frac{\partial\varphi}{\partial{t}}=A'\varphi+ \gamma B'{\varphi},
\end{equation}
\begin{equation}\label{def_A'}
\mbox{where }\ \ \ \ \ \ A'\varphi =\frac{k_BT}{\hbar}\biggl(\sum_{j=1}^{n}
\biggl(
\frac{\partial V'}{\partial x'_j} \frac{\partial\varphi}{\partial p'_j}-
  {p'_j} \frac{\partial \varphi}{\partial x'_j}
\biggr)
-{i}
\biggl(V'-\sum_{j=1}^{n}\frac{(p'_j)^2}{2} \biggr)\varphi
\biggr)
\end{equation}
\begin{equation}\label{def_B'}
\mbox{and }\ \ \ \ \  B'{\varphi}=
\sum_{j=1}^{n}\frac{\partial}{\partial{p'_j}}\left( \biggl( p'_j +i \frac{\partial}{\partial{x'_j}}\biggr){\varphi}
+\frac{\partial{\varphi} }{\partial{p'_j}}\right). \nonumber
\end{equation}

Note that the operator $A'$ is skew Hermitian, and the operator $B'$ is neither skew Hermitian neither self-adjoint.
The operator $B'$ determines the scattering process of the wave function with respect to momenta
and hence non-invertibility of the process. In this paper we consider the case when $\gamma $  is a large
quantity, i.~e. the impact of the operator $B'$ on the general evolution process of the wave function is large.
Properties of the operator $B'$ are presented in the following Theorem.

{\bf Theorem 1.} {\it The operator $B'$ given by expression (\ref{def_B'}) has a full set of eigenfunctions
(in the class of functions $\varphi [x, p]$ tending to zero at infinity) with the eigenvalues
 $0, -1, -2, \ldots.$ Respectively, the operator $B'$ is presented in the following form:
\begin{equation}\label{B_from_P}
B'=-\sum\limits_{k=0}^{\infty} k P_k,
\end{equation}
where $P_k$ are the projection operators onto the eigenspaces of the operator $B'$ with eigenvalues $-k$.

The projection operators $P_k$ satisfy the relations
\begin{equation}\label{P_property}
P_k P_k=P_k, \ \ P_k P_{k'}=0\ \ \ \mbox{for  }\ \ k\neq k',\ \ P_kB'=B'P_k=-kP_k
\end{equation}
\begin{equation}\label{E_P_property}
\mbox{and }\ \ E=\sum\limits_{k=0}^{\infty}P_k,
\end{equation}
where $E$ is the identity operator.
}

Proof of this Theorem will be given simultaneously with the proof of the following Theorem describing the form of
the projection operators $P_k$.

Denote by $H^k_{k_1...k_n}[p]\stackrel{def}{=}H_{k_1}[p_1]...H_{k_n}[p_n]$ the product of Hermite polynomials
\cite{ermit} of the corresponding variables, where
 $k=k_1+...+k_n$ is the sum of degrees of the Hermite polynomials in the product.
 By definition, the Hermite polynomial is given by the expression
 \begin{equation}\label{p_ermit}
H_{k_j}[p_j]\stackrel{def}{=}\exp\left(\frac{p^2}{2}\right)
\left(-\frac{\partial}{\partial p_j}\right)^{k_j}\exp\left(-\frac{p^2}{2}\right).
\end{equation}
Let $S_{k_1,...,k_n}$ and $I_{k_1,...,k_n}$ be the operators given by the expressions
\begin{equation}\label{S_k}
\psi_{k_1,...,k_n}=S_{k_1,...,k_n}[\varphi]\stackrel{def}{=}
\int\limits_{R^{n}}\!\! H^k_{k_1...k_n}\left[p''+i\frac{\partial}{\partial x''}\right]\varphi[x'',p''] dp'',
\end{equation}
\begin{eqnarray}\label{I_k}
\!\!\!&&\!\!\!\varphi_{k_1,...,k_n}=I_{k_1,...,k_n}[\psi_{k_1,...,k_n}]\stackrel{def}{=}\\
\!\!\!&&\!\!\!\frac{1}{(2\pi)^{3n/2}}
 \frac{1}{k_1!} ...\frac{1}{k_n!}
\int\limits_{R^{2n}}\!\!\psi_{k_1,...,k_n}[x'']H^k_{k_1...k_n}[p'-s']
e^{-\frac{(p'-s')^2}{2}} e^{i s'(x'-x'')}ds' dx''.\nonumber
\end{eqnarray}

{\bf Theorem 2.} {\it The projection operators $P_k$ have the form
\begin{eqnarray}\label{P_k}
&&P_k =\sum\limits_{\stackrel{k_1,...,k_n=0}{ k_1+...+k_n=k}}^k I_{k_1,...,k_n}S_{k_1,...,k_n},
\end{eqnarray}
and the operators $S_{k'_1,...,k'_n}$, $I_{k_1,...,k_n}$ satisfy the relations
\begin{equation}
 S_{k'_1,...,k'_n}I_{k_1,...,k_n}\psi_{k_1,...,k_n}=
\delta_{k'_1,k_1}...\delta_{k'_n,k_n}\psi_{k_1,...,k_n},
\end{equation}
where $\delta_{k'_i k_i}$ equals 0 if $k'_i \neq k_i$, and equals 1 if $k'_i =k_i$.
}

In particular, formulas (\ref{S_k}), (\ref{I_k}) and Theorem~2 imply that
\begin{eqnarray}\label{S_0}
&&\psi[x']=S_{\bar 0}[\varphi]\stackrel{def}{=}\int\limits_{R^n} \varphi[x',p'] dp',\\
\label{I_0}
&&\varphi_0[x',p']=I_{\bar 0}[\psi]\stackrel{def}{=}\frac{1}{(2\pi)^{3n/2}}
\int\limits_{R^{2n}}\!\!\psi[x'']
e^{-\frac{(p'-s')^2}{2}} e^{i s'(x'-x'')}ds' dx'',
\\
\label{P_0}
&&P_0 \varphi=\frac{1}{(2\pi)^{3n/2}}
\int\limits_{R^{3n}}\!\!\varphi[x'',p''] dp''
e^{-\frac{(p'-s')^2}{2}} e^{i s'(x'-x'')}ds' dx'',
\end{eqnarray}
\begin{eqnarray}\label{P_1}
P_1 \varphi&=&\sum\limits_{j=1}^n\frac{1}{(2\pi)^{3n/2}}
\int\limits_{R^{3n}}\!\!\left(p''_j+i\frac{\partial}{\partial x''_j}\right)\varphi[x'',p''] dp''\times\nonumber\\
&&(p'_j-s'_j)
e^{-\frac{(p'-s')^2}{2}} e^{i s'(x'-x'')}ds' dx''.
\end{eqnarray}

Note that the operators $I_{\bar 0}$ and $S_{\bar 0}$ given by formulas (\ref{I_0}) and (\ref{S_0})
make a bijection between the set of functions $\psi[x']$ and the set of eigenfunctions $\varphi_0[x',p']$
of the operator $B'$ with eigenvalue $0$.
We shall call the function $\psi[x']=S_{\bar 0}[\varphi_0[x',p']]$ by the presentation of the eigenfunction
$\varphi_0[x',p']$.

{\bf Proof of Theorems 1 and 2.}
Let us substitute into expression (\ref{def_B'}) the presentation $\varphi[x',p',t']$ in the form of the Fourier
integral with respect to $x'$ and the inverse Fourier transform:
\begin{equation} \label{fur}
  \varphi[x',p',t'] {=}\frac{1}{(2\pi)^{n/2}}
\int_{R^n}\tilde\varphi[s',p',t']e^{i s' x'}ds',
\end{equation}
\begin{equation} \label{fur_1}
\mbox{where   }\ \
  \tilde\varphi[s',p',t']\stackrel{def}{=}\frac{1}{(2\pi)^{n/2}}
\int_{R^n}\varphi[x'',p',t']e^{-{i s' x'' }}dx'',
\end{equation}
and we denoted by $s' x'$ the expression $s' x'\stackrel{def}{=}\sum_{j=1}^n s'_j x'_j$.

We obtain that the operator $B'$ has the form
\begin{equation}\label{B''}
B'[\varphi[x'',p']]= \frac{1}{(2\pi)^{n}}
\int_{R^{2n}}
\sum_{j=1}^{n}\frac{\partial}{\partial{p'_j}} \biggl(( p'_j -s'_j){\varphi}
+\frac{\partial{\varphi} }{\partial{p'_j}}\biggr)e^{i s' (x'-x'')}ds'dx''.
\end{equation}

Computing the integral over $x''$ in the left hand side of the obtained expression, taking into account equality
(\ref{fur_1}),  we obtain:

\begin{equation}\label{B'''}
B'[\varphi[x'',p']]= \frac{1}{(2\pi)^{n/2}}
\int_{R^{n}}
\sum_{j=1}^{n}\frac{\partial}{\partial{p'_j}} \biggl(( p'_j -s'_j){\tilde\varphi}
+\frac{\partial{\tilde\varphi} }{\partial{p'_j}}\biggr)e^{i s' x'}ds'.
\end{equation}
The operator
\begin{equation}\label{B_tilde}
\sum_{j=1}^{n}\frac{\partial}{\partial{p'_j}} \biggl(( p'_j -s'_j){\tilde\varphi}
+\frac{\partial{\tilde\varphi} }{\partial{p'_j}}\biggr),
\end{equation}
under the sign of integral in the previous expression, is well known (see, for example, \cite{kamke}).
This operator has a full set of eigenfunctions in the space of functions tending to zero as $|p'-s'|$ tends
to infinity.
The eigenvalues of this operator are the non-positive integers.
The eigenvalue 0 corresponds to eigenfunctions of the form
$$\tilde\varphi_0[s', p']= \tilde\psi[s'] e^{-\frac{(p'-s')^2}{2}},$$
where $\tilde\psi[s']$ is an arbitrary complex valued function of $s'\in R^n$.

The remaining eigenfunctions are obtained (as it is easy to check) by differentiation of the functions
$\tilde\varphi_0[s', p']$ with respect to $p'_j$, $j=1,\dots,n,$ and have the eigenvalues equal respectively to
$-1, -2, \ldots$, depending on the degree of the derivative.
Thus, the eigenfunctions with eigenvalues $-k=-(k_1+...+k_n)$ are the functions of the form
$$\tilde\psi_{k_1...k_n}[s'](-1)^k \frac{\partial^{k_1}}{\partial p'^{k_1}_{1}}\ldots
\frac{\partial^{k_n}}{\partial p'^{k_n}_{n}} e^{-\frac{(p'-s')^2}{2}}=
\tilde\psi_{k_1...k_n}[s']H^k_{k_1...k_n}[p'-s'] e^{-\frac{(p'-s')^2}{2}},$$
where
$\tilde\psi_{k_1...k_n}[s']$ are arbitrary complex valued functions of $s'\in R^n$, and
$H^k_{k_1...k_n}[p']=H_{k_1}[p'_1]...H_{k_n}[p'_n]$ is the product of Hermite polynomials of the corresponding
variables, and $k=k_1+...+k_n$ is the sum of degrees of the polynomials.
The Hermite polynomials of small degrees have the form
\begin{equation}
H_0=1,\ \ H_1[p_j]=p_j,\ \ H_2[p_j]=p_j^2-1.
\end{equation}

Let us represent also the functions $\tilde\psi_{k_1...k_n}[s']$ in the form of Fourier integrals:
$$
  \tilde\psi_{k_1...k_n} [s'] =\frac{1}{(2\pi)^{n/2}}
\int_{R^n}\psi_{k_1...k_n} [x''] e^{-{i s' x''}}dx''.
$$
From this formula, taking into account representation (\ref{fur})
of the function $\varphi[x',p',t']$ through $\tilde\varphi[s',p',t']$,
we obtain that the eigenfunctions
 $\varphi_{k_1...k_n}[x',p']$ of the operator $B'$ have the form
$$
\varphi_{k_1...k_n}[x',p']=\frac{1}{(2\pi)^{n}}
\int_{R^{2n}}\psi_{k_1...k_n}[x'']H^k_{k_1...k_n}[p'-s'] e^{-\frac{(p'-s')^2}{2}} e^{i s'(x'-x'')}ds' dx''.
$$

It is known \cite{ermit} that the Hermite polynomials form a complete system of functions and satisfy the
following orthogonality relations:
\begin{equation}\label{orto_ermit}
\frac{1}{(2\pi)^{n/2}} \frac{1}{k_1!}\ ...\frac{1}{k_n!}
\int_{R^{n}} H^{k'}_{k'_1...k'_n}[p'] H^k_{k_1...k_n}[p']
e^{-\frac{{p'}^2}{2}} dp'=
\delta_{k'_1 k_1} ...\delta_{k'_n k_n},
\end{equation}
where $\delta_{k'_i k_i}$ are equal to 0 if $k'_i \neq k_i$ and equal to 1 if $k'_i =k_i$
(in these formulas it is assumed that 0!=1).

This directly implies the statements of Theorems 1 and 2. In particular, the latter equality of
Theorem~1 follows from completeness of the set of eigenspaces of the operator $B'$.

Note that in representation of the eigenfunctions of the operator $B'$ in the form (\ref{I_k})
one can compute the integral over $s'$. To this end, let us substitute into this formula the expression defining
the Hermite polynomials (\ref{p_ermit}). We obtain
\begin{eqnarray}
\!\!\!&&\!\!\!\varphi_{k_1,...,k_n}=I_{k_1,...,k_n}[\psi_{k_1,...,k_n}]{=}\\
\!\!\!&&\!\!\!\frac{1}{(2\pi)^{3n/2}}
 \frac{1}{k_1!} ...\frac{1}{k_n!}\frac{(-1)^k\partial^k}{\partial {p'_1}^{k_1}...\partial {p'_n}^{k_n}}
\int\limits_{R^{2n}}\!\!\psi_{k_1,...,k_n}[x'']
e^{-\frac{(p'-s')^2}{2}} e^{i s'(x'-x'')}ds' dx''.\nonumber
\end{eqnarray}
Further, let us make the change of variables $s'=s''+p'$ under the integral, and compute the integral over
$s''$, using the well known equality that the Fourier transform of the function $\exp(-s''^2/2)$
is the function of the same form.
We obtain
\begin{eqnarray}\label{I_k_new}
\!\!\!\!\!\!\!&&\!\!\!\varphi_{k_1,...,k_n}=I_{k_1,...,k_n}[\psi_{k_1,...,k_n}]{=}\nonumber\\
\!\!\!\!\!\!\!&&\!\!\!\frac{1}{(2\pi)^{n}}
 \frac{1}{k_1!} ...\frac{1}{k_n!}\frac{(-1)^k\partial^k}{\partial {p'_1}^{k_1}...\partial {p'_n}^{k_n}}\!\!
\int\limits_{R^{n}}\!\!\psi_{k_1,...,k_n}[x'']
e^{-\frac{(x'-x'')^2}{2}} e^{i p'(x'-x'')} dx''=\nonumber\\
\!\!\!\!\!\!\!&&\!\!\!\frac{(-i)^k}{(2\pi)^{n}}
 \frac{1}{k_1!} ...\frac{1}{k_n!}\!\!
\int\limits_{R^{n}}\!\!\psi_{k_1,...,k_n}[x''] \prod_{j=1}^n (x'-x'')^{k_j}
e^{-\frac{(x'-x'')^2}{2}} e^{i p'(x'-x'')} dx''.
\end{eqnarray}
The latter equality is obtained after differentiation with respect to $p'$ required in the formula,
but under the sign of the integral.

The latter equality, in the particular case of eigenfunctions of the operator $B'$
with the zero eigenvalue, implies the following expression:
\begin{equation}\label{I_0_new}
\varphi_{0}=I_{\bar 0}[\psi_{0}]{=}
\frac{1}{(2\pi)^{n}}
\int\limits_{R^{n}}\!\!\psi_{0}[x'']
e^{-\frac{(x'-x'')^2}{2}} e^{i p'(x'-x'')} dx''.
\end{equation}

\section{The Schr\"odinger equation for the scattering process of waves and its refinement}
In the papers \cite{ben_kramers, ben_expand} the following Theorem has been proved.

{\bf Theorem 3.} {\it
The motion described by equation (\ref{eq_diff}) asymptotically splits for large $\gamma$ into
rapid motion and slow motion.

1)  After the rapid motion, an arbitrary wave function $\varphi(x, p, 0)$
goes at the time of order $1/\gamma $ to the function $\varphi_0=P_0\varphi$ which, after normalization and in the
initial coordinates, has the following form:
\begin{eqnarray}\label{view_varphi'}
\varphi_0[x, p]=\frac{1}{(2\pi{\hbar})^{n/2}}
\!\int\limits_{R^n}\!\!\psi[y,0]\chi[x-y]
e^{{{i  p(x-y)}/{\hbar}}}
dy,\\
\label{chi_def'}\mbox{where          }\ \ \ \ \ ||\psi ||=1 \ \ \ \ \mbox{and} \ \ \ \
  \chi[x-y]=\left(\frac{k_BTm}{\pi\hbar^2}\right)^{n/4}e^{-{{k_BTm}(x-y)^2}/{(2\hbar^2)}}.
\end{eqnarray}
The wave functions of kind (\ref{view_varphi'}) form the linear subspace of eigenfunctions of the operator $B$
given by (\ref{def_B}) with eigenvalue zero.
The elements of this subspace are parameterized by wave functions $\psi[y,0]=\int_{R^n}\varphi[y,p] dp$ depending only
on coordinates $y\in R^n$.

2) The slow motion starting with a wave function $\varphi_0[x, p]$  of the form (\ref{view_varphi'}) with nonzero
function $\psi[y,0]$,   goes along the subspace of such functions and is parameterized by the wave function
$\psi[y,t]$ depending on time. The function $\psi[y,t]$ satisfies the following Schr\"odinger equation:
$ i\hbar {\partial \psi}/{\partial t} = \hat{H}\psi$,
where the action of the operator
$\hat{H}$ for $\gamma \rightarrow \infty$ has the form
\begin{equation}\label{hatH}
{\hat H}\psi =- \frac{\hbar^2}{2m}\biggl(\sum_{k=1}^{n}\frac{\partial^2 \psi }{\partial{y^2_k}}\biggr)
+V[y] \psi-\frac{k_BT}{2}n\psi.
\end{equation}}
Proof of the first part of Theorem~3 is given in \cite{ben_expand}.
Proof of the second part of this Theorem is given in \cite{ben_kramers},
however, in that paper there is an error in the sign before $({k_BT}/{2})n\psi$.

Theorem 3 describes a solution of equation (\ref{eq_diff}) and respectively
of equation (\ref{eq_diff'})   in the zero deterministic approximation in the parameter $1/\gamma$ after the
transitional process in time of order $1/\gamma$. The purpose of the present paper is proof
of a Theorem refining the result of the part of Theorem~3 describing the slow motion.
To this end, we construct the next approximation of equation (\ref{eq_diff'}) with respect to the parameter $1/\gamma$.

Thus, let us pass to approximate description of the generalized Kramers equation (\ref{eq_diff'}) for large $\gamma$
by means of systematic decomposition over powers of $\gamma^{-1}$.
The method used here is similar to the method given in the book by van Kampen \cite{van_kampen}.
In this book, from the Kramers equation describing the Brownian motion of a particle in the phase space,
one deduces the Fokker--Planck equation describing approximately the same process,
but in the form of the Brownian motion of the particle in configuration space after certain time of
transitional process. During this process the distribution with respect to momenta becomes the Maxwell distribution.

Let us rewrite equation (\ref{eq_diff'}) in the following form:
\begin{equation}\label{vankampen_diff}
B'\varphi=\frac{1}{\gamma}\left(\frac{\partial\varphi}{\partial t} -A'\varphi\right).
\end{equation}
Let us look for solution of this equation in the form
\begin{equation}\label{varphi_gamma}
\varphi=\varphi_0+\gamma^{-1}\varphi_1+\gamma^{-2}\varphi_2+...
\end{equation}
Let us substitute expression (\ref{varphi_gamma}) into equation (\ref{vankampen_diff}), and write out the equations
for coefficients before equal powers of $\gamma^{-1}$. We obtain:
\begin{eqnarray}\label{eq_gamma_0}
\mbox{for }\gamma^0:\ \ &\ \ \ \ &B'\varphi_0=0;\\
\label{eq_gamma_1}
\mbox{for }\gamma^{-1}:&\ \ \ \ &B'\varphi_1=\frac{\partial\varphi_0}{\partial t} -A'\varphi_0;
\\
\label{eq_gamma_2}
\mbox{for }\gamma^{-2}:&\ \ \ \ &B'\varphi_2=\frac{\partial\varphi_1}{\partial t} -A'\varphi_1; \ldots.
\end{eqnarray}
Equation (\ref{eq_gamma_0}) implies that $\varphi_0$ belongs to the subspace of eigenfunctions of the operator $B'$
with eigenvalue 0, i.~e. $\varphi_0=P_0\varphi_0$, where $P_0$ is the projection operator onto the eigenspace
of the operator $B'$ with eigenvalue 0.

Let us apply to both parts of equality (\ref{eq_gamma_1}) the projection operator $P_0$ from the left.
Taking into account equalities
$P_0 B'=0 $ and $\varphi_0=P_0\varphi_0$, we obtain:
\begin{equation}\label{cond1}
0=\frac{\partial\varphi_0}{\partial t} -P_0A'P_0\varphi_0.
\end{equation}
(Note that the operator $P_0A'P_0$  corresponds to the Schr\"odinger operator $H'$ in Theorem~3.)

Provided equality (\ref{cond1}) holds, equation (\ref{eq_gamma_1}) has a solution which we represent in the form
\begin{eqnarray}\label{f_1}
\varphi_1=\varphi_{1,0}+f_1,\ \ \mbox{where}\ P_0 f_1=0,\ \ \varphi_{1,0}= P_0\varphi_{1}=P_0\varphi_{1,0},\nonumber\\
f_1=B^{'-1}(P_0A'P_0\varphi_0-A'P_0\varphi_0),
\end{eqnarray}
and $B^{'-1}$ is the inverse operator to $B'$ on the subspace spanned by eigenfunctions of the operator $B'$
with nonzero eigenvalues. Formula (\ref{B_from_P}) for $B'$ implies that the operator $B^{'-1}$  has the form
\begin{equation}\label{invertB'}
B^{'-1}=-\sum_{k=1}^\infty k^{-1}P_k.
\end{equation}
This and the properties (\ref{P_property}) of projection operators imply that $P_0 B^{'-1}=0$ and
$P_k B^{'-1}=-k^{-1}P_k$.

By the equality $B^{'-1}P_0=0$ the expression for $f_1$ in formula (\ref{f_1}) takes the form
\begin{eqnarray}\label{f_1_new}
f_1=-B^{'-1}A'P_0\varphi_0.
\end{eqnarray}
Let us substitute the expression $\varphi_1=\varphi_{1,0}+f_1$ from formula (\ref{f_1}) into equality (\ref{eq_gamma_2}),
and apply to both parts of equality (\ref{eq_gamma_2}) the projection operator $P_0$ from the left.  Using the equalities
$P_0 B'=0$, $P_0 f_1=0$, and $\varphi_{1,0}= P_0\varphi_{1,0}$, and equality (\ref{f_1_new}), after the substitutions
mentioned above and opening the brackets we obtain:
$$0=P_0\left(\frac{\partial\varphi_1}{\partial t} -A'\varphi_1\right)\ \ \mbox{or}$$
\begin{equation}\label{cond2}
0=\frac{\partial\varphi_{1,0}}{\partial t} -P_0A'P_0\varphi_{1,0}+P_0 A'B^{'-1}A'P_0\varphi_0.
\end{equation}
Let us now sum up equations (\ref{cond1}) and (\ref{cond2}) multiplied respectively by $1$ and $\gamma^{-1}$.
Then for the function $\varphi_{\gamma 1,0}$, defined by the equality \begin{equation}\label{varphi_gamma 1_0}
\varphi_{\gamma 1,0}\stackrel{def}{=}\varphi_0+\gamma^{-1}\varphi_{1,0},
\end{equation}
 we obtain the following equation up to summands of order $\gamma^{-1}$:
\begin{eqnarray}\label{eq_diff'_gamma}
0\!\!\!&=&\!\!\!
\frac{\partial\varphi_{\gamma 1,0}}{\partial t} -P_0A'P_0\varphi_{\gamma 1,0}
+\gamma^{-1}P_0 A'B^{'-1}A'P_0\varphi_{\gamma 1,0}
+O[\gamma^{-2}].
\end{eqnarray}
Respectively, for function $\varphi_{\gamma 1}\stackrel{def}{=}\varphi_0+\gamma^{-1} \varphi_1$,
which takes into account only first two summands in the decomposition (\ref{varphi_gamma}) for $\varphi$,
using the equality $\varphi_1=\varphi_{1,0}+f_1$ in expression (\ref{f_1}) and substituting into
it expression (\ref{f_1_new}) for $f_1$,  we obtain the following equality up to summands of order $\gamma^{-1}$:
\begin{eqnarray}\label{varphi_gamma_leq1}
\varphi_{\gamma 1}&=&
\varphi_{\gamma 1,0}
-\gamma^{-1}B^{'-1}A'P_0\varphi_{\gamma 1,0}
+O[\gamma^{-2}].
\end{eqnarray}

Since $\varphi_{\gamma 1,0}=P_0 \varphi_{\gamma 1,0}$, that is, $\varphi_{\gamma 1,0}$ is an eigenfunction
of the operator $B'$ with eigenvalue 0, then, in accordance with formulas (\ref{S_0}) and (\ref{I_0})
and Theorem~2, this function is completely determined by the function
$\psi\stackrel{def}{=}S_{\bar 0} [\varphi_{\gamma 1,0}]$
by the formula
$\varphi_{\gamma 1,0}=I_{\bar 0} [\psi]$, where $S_{\bar 0}$ and $I_{\bar 0}$ are the operators defined by formulas
(\ref{S_0}) and~(\ref{I_0}).
As above, let us call the function $\psi$ by the presentation of the eigenfunction $\varphi_{\gamma 1,0}$.

In order to obtain the $\psi$-presentation of equation (\ref{eq_diff'_gamma}), let us substitute into it,
instead of function $\varphi_{\gamma 1,0}$ the equal expression $I_{\bar 0} [\psi]$, let us act on both
parts of equation by the operator $S_{\bar 0}$, and let us use the following equalities deduced from
relations of Theorem~2:
$$P_0\varphi_{\gamma 1,0}=I_{\bar 0}S_{\bar 0}[\varphi_{\gamma 1,0}]=I_{\bar 0}[\psi],\ \
S_{\bar 0}P_0= S_{\bar 0},\ \ \ \mbox{and}\ \ \psi=S_{\bar 0}I_{\bar 0}[\psi].$$
We obtain:
\begin{eqnarray}\label{eq_diff'psi_gamma}
0&=&
\frac{\partial\psi}{\partial t} -S_{\bar 0}A'I_{\bar 0}\psi
+\gamma^{-1}S_{\bar 0} A'B^{'-1}A'I_{\bar 0}\psi
+O[\gamma^{-2}].
\end{eqnarray}

Respectively, expression (\ref{varphi_gamma_leq1}) for function $\varphi_{\gamma 1}[x',p']$, presented through
the function $\psi[x']$, without account of summands of order $\gamma^{-2}$, reads
\begin{eqnarray}\label{varphi_gamma_1_psi}
\varphi_{\gamma 1}&=&
I_{\bar 0}[\psi]
-\gamma^{-1}B^{'-1}A'I_{\bar 0}[\psi]
+O[\gamma^{-2}].
\end{eqnarray}

Note that this equality implies $S_{\bar 0}\varphi_{\gamma 1}=\psi$. Thus, expression (\ref{varphi_gamma_1_psi})
and the operator $S_{\bar 0}$ yield mutually inverse bijections between the set of functions $\varphi_{\gamma 1}[x',p']$
and the set of functions $\psi[x']$. Thus, function $\psi[x']$ is also a presentation of the function
$\varphi_{\gamma 1}[x',p']$ by formula (\ref{varphi_gamma_1_psi}). And the function $\psi[x']$ evolves in time
according to equation~(\ref{eq_diff'psi_gamma}).

 Thus, from equation (\ref{eq_diff'psi_gamma}) and relation $P_0=I_{\bar 0}S_{\bar 0}$ of Theorem~2 we obtain
 the following approximate equation, with account of summands up to order $\gamma^{-1}$, for the slow
 subprocess in the process described by the modified Kramers equation~(\ref{eq_diff'}):
\begin{eqnarray}\label{eq_diff'_1_psi}
\frac{\partial\psi}{\partial t}\!\!\!\!&=&\!\!\!\!
S_{\bar 0}A'I_{\bar 0}\psi
-\gamma^{-1}S_{\bar 0} A'B^{'-1}A'I_{\bar 0}\psi
+O[\gamma^{-2}].
\end{eqnarray}
(The form of the first summand in the right hand side of the equation, namely, the operator $S_{\bar 0}A'I_{\bar 0}$,
is known to us from Theorem~3.) The complete description of the right hand side of this equation is given by the following
Theorem.

{\bf Theorem 4.} {\it
 The slow motion mentioned in Theorem~3, starting from the wave function $\varphi_0[x, p]$ of the form (\ref{view_varphi'})
 with nonzero function $\psi[y,0]$, and parameterized by the wave function $\psi[y,t]$,
 satisfies the modified Schr\"odinger equation of the form
$ i\hbar {\partial \psi}/{\partial t} = \hat{H_1}\psi$,  where action of the operator $\hat{H_1}$ is the following:
\begin{eqnarray}\label{hatH_1}
{\hat H_1}\psi\!& =&\!- \frac{\hbar^2}{2m}\sum_{k=1}^{n}\!\frac{\partial^2 \psi }{\partial{y^2_k}}
+V \psi-\frac{k_BTn}{2}\psi +\frac{i\gamma^{-1}}{4}
\biggl(\sum_{j=1}^n\!\frac{\hbar}{m} \frac{\partial^2 V}{\partial y_j^2} -\frac{(k_BT)^2n}{\hbar}\biggr)\psi+\nonumber\\
&&+O[\gamma^{-2}].
\end{eqnarray}
}

Proof of Theorem 4 is given in the Appendix.

The constants in operators $\hat H$ and $\hat H_1$ of Theorems~3 and~4 can be neglected, they are inessential.
Using the standard method of perturbation theory \cite{landau},
let us compute the corrections to the eigenvalues and eigenfunctions of the Hamilton operator for the operator $\hat H_1$.

Let $E_n^{(0)}$ be the eigenvalues of the Hamilton operator $\hat H$, and $\psi_n^{(0)}$ be the corresponding
eigenfunctions. Let $E_n$ and $\psi_n$ be the eigenvalues and eigenfunctions of the operator $\hat H_1$.
Then by definition of eigenfunctions, one has the equalities
$\hat H\psi_n^{(0)}=E_n^{(0)}\psi_n^{(0)}$ and
$\hat H_1\psi_n=E_n\psi_n$. Let us look for $E_n$ and $\psi_n$
 in the form
 \begin{eqnarray}
E_n&=&E_n^{(0)}+\gamma^{-1}E_n^{(1)}+O(\gamma^{-2})\\
\label{psi_n}
\psi_n&=&\psi_n^{(0)}+\gamma^{-1}\sum\limits_{\stackrel{m}{m\neq n}}c_{nm}\psi_m^{(0)}+O(\gamma^{-2}).
\end{eqnarray}
Let us substitute these expressions into equality (\ref{hatH_1}).
In the obtained expression, let us equate coefficients before the corresponding powers of $\gamma^{-1}$, and take
the scalar products of both parts of the obtained equalities with $\psi_n^{(0)}$ or $\psi_k^{(0)}$.
Using the orthonormality of the system of eigenfunctions $\psi_k^{(0)}$ with respect to the scalar
product $\langle\ ;\ \rangle$ given by
\begin{equation}
\langle\psi_k;\psi_m \rangle=
 \int\limits_{R^{n}}\psi_k[y]\psi_k^{*}[y]dx,
\end{equation}
 we obtain:
\begin{eqnarray}
E_n^{(1)}&=&\frac{i\hbar}{4m}\langle\Delta^2V\psi_n^{(0)}; \psi_n^{(0)}\rangle\\
\label{c_nk}
c_{nk}&=&\frac{i\hbar}{4m}\frac{\langle\Delta^2V\psi_n^{(0)}; \psi_k^{(0)}\rangle} {E_n^{(0)}-E_k^{(0)}},\\
\mbox{where}\ \ \ \Delta^2V&=&\sum_{k=1}^{n}\!\frac{\partial^2 V }{\partial{y^2_k}}.
\end{eqnarray}
These equalities and the modified Schr\"odinger equation $ i\hbar {\partial \psi}/{\partial t} = \hat{H_1}\psi$
imply that the eigenfunction $\psi_n$ evolves in time according to the following expression:
$$\psi_n[t]=\psi_n [0] \exp\left(-\frac{iE_n t}{\hbar}\right)=
\psi_n [0]  \exp\left(-\frac{iE_n^{(0)} t}{\hbar}-\frac{i\gamma^{-1}E_n^{(1)} t}{\hbar}\right),
$$
$$\mbox{where}\ \ \ -\frac{i\gamma^{-1}E_n^{(1)}}{\hbar}=\frac{\gamma^{-1}}{4m}\langle\Delta^2V\psi_n^{(0)};
\psi_n^{(0)}\rangle\ \ \ \mbox{is real.}$$
Therefore, the absolute value of the eigenfunction $\psi_n$ changes in time exponentially with the
exponent
$({\gamma^{-1}}/{4m})\langle\Delta^2V\psi_n^{(0)}; \psi_n^{(0)}\rangle t. $
Thus, if the system at the initial moment of time is in the state $\psi[0]$, where
$\psi [0]=\sum_n^\infty a_n \psi_n$ is a certain superposition of eigenstates of the operator $\hat H_1$,
then after the time $t$  the system will be in the state $\psi[t]/|\psi[t]|$,  where
$\psi [t]=\sum_n^\infty a_n \psi_n[t]$. Since the absolute values of the eigenstates $\psi_n[t]$
change exponentially with different velocities, then after large enough time the state
 $\psi[t]/|\psi[t]|$ will be close to certain eigenstate $\psi_k$, for which the value $-iE_k^{(1)}$
 is maximal among the values $-iE_n^{(1)}$ for $n$ with nonzero $a_n$ in the sum $\sum_n^\infty a_n \psi_n$.
 This phenomenon has been called decoherence and has been studied in a series of papers \cite{zeh, zurek, mensky}.
 The time of decoherence in our model can be estimated as the time $t$ for which
 $a_k\exp(-iE_k^{(1)t/\hbar})>>a_{k_1}\exp(-iE_{k_1}^{(1)}t/\hbar)$, where
$-iE_{k_1}^{(1)}$ is the next value after the maximal value $-iE_{k}^{(1)}$ among the numbers $-iE_n^{(1)}$ for
$n$ with nonzero $a_n$ in the sum $\sum_n^\infty a_n \psi_n$.

Besides that, formulas (\ref{psi_n}) and (\ref{c_nk}) imply that the eigenfunctions $\psi_n$ of the operator $\hat H_1$
are in general case not orthogonal to each other. We have:
$$\langle \psi_n;\psi_k\rangle=\gamma^{-1}(c_{nk}+c^*_{kn})=2\gamma^{-1}c_{nk}.
$$

Let $\varphi_n[x,p]$ and $\varphi_k[x,p]$ be wave functions on the phase space corresponding to functions
$\psi_n$ and $\psi_k$
by formula (\ref{varphi_gamma_1_psi}).
 According to the assumptions of the model, the square of the absolute value of the scalar product
 of normalized functions
$\varphi_n[x,p]$ and $\varphi_k[x,p]$  yields the probability to find the system in the state $\psi_k$,
if it is in the state $\psi_n$. The latter equality and formula (\ref{varphi_gamma_1_psi})
imply that this probability is nonzero if $c_{nk}\neq 0$.

\section{Conclusion}
In this paper we have constructed an approximate description of the slow phase of the scattering process
of the wave function up to $\gamma^{-1}$,
where $\gamma$ is the resistance of the medium per unit of mass of a particle--wave.
The obtained approximation is described by the Schr\"odinger equation, supplemented with a summand with
coefficient $\gamma^{-1}$.
In this approximation one has effects of decoherence and spontaneous jumps from one level to another.

Note that for a free particle, when $V=0$, and for harmonic oscillator, when the second derivatives
of the potential are constant,
the summand of the operator $\hat H_1$ with the factor $\gamma^{-1}$ in Theorem~4 is a constant.
Therefore, in the approximation up to $\gamma^{-1}$, due to this summand, in these cases all wave functions
decrease in amplitude with the same velocity, and dissipation is non-observable.
Hence, to take into account dissipation in this model either for free particle or for harmonic oscillator,
one should consider the summands with factors $\gamma^{-2}$ and $\gamma^{-3}$.
Note also that the method of construction of the $\gamma^{-1}$ summand in the modified Schr\"odinger equation
used in this paper, allows, in principle, to construct also the summands with factors $\gamma^{-2}$ and $\gamma^{-3}$.

The next phenomenon which can appear in this model, is the occurrence of nonzero width of spectral lines.
The interaction of a quantum particle with medium should cause occurrence of width of energy levels for the energy
operator.
That is, if $\psi_j$ is an eigenfunction of the Hamilton operator with eigenvalue $E_j$,
then the corresponding wave function $\varphi_j[x',p']$ in the phase space
is given by the formula~(\ref{varphi_gamma_1_psi}).
According to the assumption of the model, the function $\varphi_j[x',p']$ defines the density function
of probability distribution in the phase space in the form $\varphi_j[x',p'] \varphi^*_j[x',p']$. Respectively,
the average value of the energy function $H[x',p']$ in this case is given by the following expression:
\begin{equation}
 E_j=\int\limits_{R^{2n}}H[x',p'] \varphi_j[x',p'] \varphi^*_j[x',p'] dx'dp'.
\end{equation}
Then $(\Delta E_j)^2$, the average of square of deviation from the average energy, is computed by the formula
\begin{equation}
(\Delta E_j)^2=\int\limits_{R^{2n}}(H[x',p']-E_j)^2 \varphi_j[x',p'] \varphi^*_j[x',p'] dx'dp'.
\end{equation}

These computations can be completed for concrete quantum systems, to obtain the dependence of
$\Delta E_j$ on $T$ and $\gamma$.
After that, the obtained data can be compared with the experimental data on the width of spectral lines.

\subsection*{Appendix.  Proof of Theorem 4}

 To prove Theorem 4 one should compute the right hand side of equation (\ref{eq_diff'_1_psi}),
 and pass to the initial coordinates.

The first summand in the right hand side of this equation is computed in \cite{ben_kramers}
(as we said above, in this paper the sign before the last summand in this formula has been computed incorrectly).
It has the form of the Schr\"odinger operator, which in new coordinates (\ref{new_var}) reads as follows:
$$
S_{\bar 0}A'I_{\bar 0}\psi = -i \frac{k_BT}{\hbar} \left(
-\frac{1}{2}\sum_{j=1}^{n}\frac{\partial^2}{\partial (x'_j)^2}+V'- \frac{n}{2} \right)\psi.
$$
We are going to compute the operator $S_{\bar 0}A' B^{'-1}A'I_{\bar 0}$ which stands in equation (\ref{eq_diff'_1_psi})
with the factor $\gamma^{-1}$.

Let us first transform this expression using the equality
$$B^{'-1}=-\sum_{k=1}^\infty k^{-1}P_k$$ 
obtained above (\ref{invertB'}), 
 the relations 
$P_k=\sum_{k_1+ ...+k_n=k}I_{k_1, ...,k_n}S_{k_1, ...,k_n}$ from Theorem~2, and
  $P_k P_0=0$ for $k>0$ .

We obtain:
\begin{eqnarray} \label{S_0AB}
S_{\bar 0}A' B^{'-1}A'I_{\bar 0}
&=&
-\sum_{k=1}^\infty k^{-1} S_{\bar 0}A' P_k A'I_{\bar 0}=
\nonumber\\
&=&
-\!\!\!\!\!\!\!\!\!\!\!
\sum_{\stackrel {k_1, ...,k_n=0}{k_1+ ...+k_n=k\geq1}}^\infty\!\!\!\!\!\!\!\!\!\!\!
k^{-1} S_{\bar 0}A'I_{k_1, ...,k_n}S_{k_1, ...,k_n} A'I_{\bar 0}
\end{eqnarray}

Let us now compute the operator
$A'I_{k_1,...,k_n}$, where the operator $A'$ is given by expression (\ref{def_A'}),
and the operator $I_{k_1,...,k_n}$  is defined by expression~(\ref{I_k}).

We have
\begin{eqnarray}\label{op1}
A'I_{k_1,...,k_n}\psi\!\!\!\!&=&\!\!\!\!
\frac{k_BT}{\hbar}\biggl(\sum_{j=1}^{n}
\biggl(
\frac{\partial V'}{\partial x'_j} \frac{\partial}{\partial p'_j}-
  {p'_j} \frac{\partial }{\partial x'_j}
\biggr)
-{i}
\biggl(V'-\sum_{j=1}^{n}\frac{p^{'2}_j}{2} \biggr)
\biggr)\circ
\\
\!\!\!\!\!\!\!\!&&\!\!\!\!\!\!\!\!\!\!\!\!\!\!\!\!
\frac{1}{(2\pi)^{3n/2}}
 \frac{1}{k_1!} ...\frac{1}{k_n!}
\int\limits_{R^{2n}}\!\!\psi[x'']H^k_{k_1...k_n}[p'-s']
e^{-\frac{(p'-s')^2}{2}} e^{i s'(x'-x'')}ds' dx''.\nonumber
\end{eqnarray}

Let us put the summands of the operator $A'$ in this expression under the sign of integral,
and let us decompose the functions of $(p'-s')$ under the sign of integral over Hermite polynomials
$H^l_{l_1...l_n}[p'-s']$. Using the equalities
\begin{eqnarray}
 \!\!\!\!\!\!\!\!\!\!\!\!&&\!\!\!\!\!\!\!\!\frac{\partial}{\partial p'_j}\left(H^k_{k_1...k_n}[p'-s']
 e^{-\frac{(p'-s')^2}{2}}\right)\stackrel{def}{=}-H^{k+1}_{k_1...k_j+1...k_n}[p'-s'] e^{-\frac{(p'-s')^2}{2}},\\
\!\!\!\!\!\!\!\!\!\!\!\!&&\!\!\!\!\!\!\!\!\frac{\partial }{\partial x'_j}e^{i s'(x'-x'')}=is'_j e^{i s'(x'-x'')},\\
\!\!\!\!\!\!\!\!\!\!\!\!&&\!\!\!\!\!\!\!\! p_j^{'2}=((p'_j-s'_j)^2-1)+2p'_j s'_j-s_j^{'2}+1=H_2[p'_j-s'_j]
+2p'_j s'_j-s_j^{'2}+1,
\end{eqnarray}
we obtain:
\begin{eqnarray}
A'I_{k_1,...,k_n}\psi\!\!\!\!&=&\!\!\!\!
\frac{k_BT}{\hbar}\frac{1}{(2\pi)^{3n/2}}
 \frac{1}{k_1!} ...\frac{1}{k_n!}
\int\limits_{R^{2n}}
\biggl(-\sum_{j=1}^{n}
\frac{\partial V'}{\partial x'_j}H^{k+1}_{k_1...k_j+1...k_n}[p'-s']
\nonumber\\
&&
-i\sum_{j=1}^{n}p'_j s'_j H^{k}_{k_1...k_n}[p'-s']
-iV'[x']H^{k}_{k_1...k_n}[p'-s']
\nonumber\\
&&
+\frac{i}{2}\sum_{j=1}^{n}(H_2[p'_j-s'_j]+2p'_j s'_j-s_j^{'2}+1)
H^{k}_{k_1...k_n}[p'-s']
\biggr)
\nonumber\\
&&
\times\psi[x''] e^{-\frac{(p'-s')^2}{2}} e^{i s'(x'-x'')}ds' dx''.
\end{eqnarray}
After opening the brackets in the latter summand and summing up the terms similar to the second summand,
we obtain:
\begin{eqnarray}\label{A'I_k}
A'I_{k_1,...,k_n}\psi\!\!\!\!&=&\!\!\!\!
\frac{k_BT}{\hbar}\frac{1}{(2\pi)^{3n/2}}
 \frac{1}{k_1!} ...\frac{1}{k_n!}
\int\limits_{R^{2n}}
\biggl(-\sum_{j=1}^{n}
\frac{\partial V'}{\partial x'_j}H^{k+1}_{k_1...k_j+1...k_n}[p'-s']
\nonumber\\
&&\!\!\!\!\!\!\!\!\!\!\!\!\!\!\!\!\!\!
-iV'[x']H^{k}_{k_1...k_n}[p'-s']\!+\!
\frac{i}{2}\sum_{j=1}^{n}H_2[p'_j-s'_j]H^{k}_{k_1...k_n}[p'-s']
\nonumber\\
&&\!\!\!\!\!\!\!\!\!\!\!\!\!\!\!\!\!\!
-\frac{i}{2}\sum_{j=1}^{n}(s_j^{'2}-1)
H^{k}_{k_1...k_n}[p'-s']\!
\biggr)\psi[x'']
 e^{-\frac{(p'-s')^2}{2}} e^{i s'(x'-x'')}ds' dx''.
\end{eqnarray}

{\bf Lemma 1.} {\it  The operators $S_{\bar 0}A'I_{k_1,...,k_n}$ read as follows:
\begin{equation}
S_{\bar 0}A'I_{\bar 0}[\psi]=\frac{k_B T}{\hbar}\biggl(-iV'\psi
 +\frac{i}{2}\sum_{j=1}^n \frac{\partial^2 \psi}{\partial x^{'2}_j}+
\frac{i n}{2}\psi
\biggr);
\end{equation}
\begin{eqnarray}\label{S_0A'I_k}
&&S_{\bar 0}A'I_{k_1,...,k_n}[\psi]=\frac{i}{2}\frac{k_B T}{\hbar}
\psi, \ \ \mbox{when}\ \ k= k_1+...+k_n=2\nonumber\\
&& \mbox{and only one of }\ \ {k_1,...,k_n}\ \ \ \mbox{equals 2};
\end{eqnarray}
In the remaining cases $S_{\bar 0}A'I_{k_1,...,k_n}=0.$
}

{\bf Proof.} According to formula ({\ref{S_0}) the operator $S_{\bar 0}$ is an integration over $p$.
Hence for proof of the formulas of Lemma~1 one should compute integral over $p$
of expression~(\ref{A'I_k}) for the operator $A'I_{k_1,...,k_n}$.
To this end, let us use the orthogonality formula (\ref{orto_ermit}) for Hermite polynomials and the fact that
 $H^0_{\bar 0}=1$. This implies that integral over $p$ of the first summand of expression (\ref{A'I_k}) equals~0.
 Integral of $H^{k}_{k_1...k_n}[p'-s']$  in the second and the fourth summand with respect to
$1/(2\pi)^{(n/2)}\exp[-(p'-s')^2/{2}]dp$ equals 1 only for $k=0$,  and in the other cases also equals~0.
Integral of $H_2[p'_j-s'_j]H^{k}_{k_1...k_n}[p'-s']$  in the third summand with respect to
$1/(2\pi)^{(n/2)}\exp[-(p'-s')^2/{2}]dp$ differs from 0 and equals 2 only when  $k=2$, $k_j=2$,  and the rest of
$k_1, ..., k_n$  equal~0.
Then in the obtained expression we compute integrals over $s'$ and $x''$, using the fact that integral of
${1}/{(2\pi)^{n}} e^{i s'(x'-x'')}$ over $s'$ is the delta function at the point $x'$, and integral of
expression ${1}/{(2\pi)^{n}}{s'}^{2} e^{i s'(x'-x'')}$ over $s'$ is the delta function at the point $x'$
with $-\partial^2/\partial {x''}^{2}$.
As a result of these computations we get the statement of Lemma~1.

Since by Lemma 1 the operators $S_{\bar 0}A'I_{k_1,...,k_n}$  with $k= k_1+...+k_n>0$ are not equal to 0
only when $k=2, k_j=2$ (and therefore the remaining $k_1, ..., k_n$  equal 0),
then for computation of operator $S_{\bar 0}A' B^{'-1}A'I_{\bar 0}$ by formula (\ref{S_0AB}) it remains to
compute the operators $S_{k_1,...,k_n}A'I_{\bar 0}$ for the same values of $k_1, ..., k_n$.

{\bf Lemma 2.} {\it  The operators $S_{k_1,...,k_n}A'I_{\bar 0}$, in the case when $k_j=2$ and $k_1+...+k_n=2$,
read as follows:
\begin{equation}
S_{k_1,...,k_n}A'I_{\bar 0}[\psi]=\frac{k_B T}{\hbar}\biggl(
-i\frac{\partial^2 V'}{\partial x^{'2}_j}\psi+i\psi
\biggr).
\end{equation}
}

{\bf Proof.} Under the conditions of Lemma 2, when $k=k_j=2$, the operator $S_{k_1,...,k_n}$ reads, by formula (\ref{S_k}),
as follows:
\begin{eqnarray}
S_{k_1,...,k_n}[\varphi]&\stackrel{def}{=}&\int\limits_{R^{n}}\!\!
H^k_{k_1...k_n}\left[p'+i\frac{\partial}{\partial x'}\right]\varphi[x',p'] dp'
\nonumber\\
&& =\int \limits_{R^{n}} \!\! H_2\left(p'_j+i\frac{\partial}{\partial x'_j}\right)
\varphi[x',p'] dp'.
\end{eqnarray}
This and formula (\ref{A'I_k}) for $A'I_{\bar 0}$ imply that
\begin{eqnarray}
S_{k_1,...,k_n}A'I_{\bar 0}[\psi]
\!\!\!\!\!\!\!\!&&=\frac{k_BT}{\hbar}\frac{1}{(2\pi)^{3n/2}}
\int\limits_{R^{3n}}\!\!H_2\left(p'_j+i\frac{\partial}{\partial x'_j}\right)\circ
\nonumber\\
&&
\circ\biggl(-\sum_{j'=1}^{n}
\frac{\partial V'[x']}{\partial x'_{j'}}H_1[p'_{j'}-s'_{j'}]-
iV'[x']\!+\!
\frac{i}{2}\sum_{j'=1}^{n}H_2[p'_{j'}-s'_{j'}]
\nonumber\\
&&
-
\frac{i}{2}\sum_{j'=1}^{n}({s'}_{j'}^{2}-1)
\!
\biggr)
\psi[x'']
 e^{-\frac{(p'-s')^2}{2}} e^{i s'(x'-x'')}ds' dx'' dp'.
\end{eqnarray}
Taking into account that by definition of Hermite polynomials we have
$$H_2\left(p'_j+i\frac{\partial}{\partial x'_j}\right)=
\left(\left(p'_j+i\frac{\partial}{\partial x'_j}\right)^2
-1\right),
$$
let us apply this operator in the previous expression, using the formula for the derivative ${\partial}/{\partial x'_j}$
of a product of functions depending on $x'$, and the equality
$$ i\frac{\partial}{\partial x'_j}e^{i s'(x'-x'')}=-s'e^{i s'(x'-x'')}.$$
We obtain:
\begin{eqnarray}
S_{k_1,...,k_n}A'I_{\bar 0}[\psi]
\!\!\!\!\!\!\!\!&&=\frac{k_BT}{\hbar}\frac{1}{(2\pi)^{3n/2}}
\int\limits_{R^{3n}}\!\!H_2\left(p'_j-s'_j\right)
\nonumber\\
&&\!\!\!\!\!\!\!\!\!\!\!\!\!\!\!\!
\times\biggl(-\sum_{j'=1}^{n}
\frac{\partial V'[x']}{\partial x'_{j'}}H_1[p'_{j'}-s'_{j'}]-
iV'[x']\!+\!
\frac{i}{2}\sum_{j'=1}^{n}H_2[p'_{j'}-s'_{j'}]
\nonumber\\
&&
-
\frac{i}{2}\sum_{j'=1}^{n}({s'}_{j'}^{2}-1)
\!
\biggr)
\psi[x'']
 e^{-\frac{(p'-s')^2}{2}} e^{i s'(x'-x'')}ds' dx'' dp'
\\
&&\!\!\!\!\!\!\!\!\!\!\!\!\!\!\!\!
+\frac{k_BT}{\hbar}\frac{1}{(2\pi)^{3n/2}}
 \int\limits_{R^{3n}}\!\!\biggl(
-2i(p'_j-s'_j)\sum_{j'=1}^{n}
\frac{\partial^2 V'[x']}{\partial x'_{j}\partial x'_{j'}}H_1[p'_{j'}-s'_{j'}]
\nonumber\\
&&\!\!\!\!\!\!\!\!\!\!\!\!\!\!\!\!
+\sum_{j'=1}^{n}
\frac{\partial^3 V'[x']}{\partial^2 x'_{j}\partial x'_{j'}}H_1[p'_{j'}-s'_{j'}]
+2(p'_j-s'_j)\frac{\partial V'[x']}{\partial x'_{j}}
+i\frac{\partial^2 V'[x']}{\partial{x'_{j}}^2}
\biggr)
\nonumber\\
&&\ \ \ \ \ \ \ \ \ \ \ \ \ \ \ \ \ \ \ \
\times
\psi[x'']
 e^{-\frac{(p'-s')^2}{2}} e^{i s'(x'-x'')}ds' dx'' dp'.
\end{eqnarray}

Further, let us integrate over $p'$ in each integral in the obtained expression, using the
orthogonality relation (\ref{orto_ermit}) for Hermite polynomials and the equalities $1=H_0(p'_j-s'_j)$ and
$(p'_j-s'_j)=H_1(p'_j-s'_j)$. We obtain:
\begin{eqnarray}
&&\!\!\!\!\!\!\!\!\!\!S_{k_1,...,k_n}A'I_{\bar 0}[\psi]
=\frac{k_BT}{\hbar}\frac{1}{(2\pi)^{n}}
\int\limits_{R^{2n}}\!\!\left(\!-0-0+\frac{i}{2}2-0\right)\!\!
\psi[x'']
  e^{i s'(x'-x'')}ds' dx''
\nonumber\\
&&\!\!\!\!\!\!\!\!\!\!\!
+\frac{k_BT}{\hbar}\frac{1}{(2\pi)^{n}}\!
 \int\limits_{R^{2n}}\!\!\!\left(\!\!
-2i
\frac{\partial^2 V'[x']}{\partial {x'_{j}}^2}+0+0
+i\frac{\partial^2 V'[x']}{\partial{x'_{j}}^2}
\right)\!\!
\psi[x'']
  e^{i s'(x'-x'')}ds' dx''.
\end{eqnarray}

After computing the integrals over $s'$ and $x''$ in the obtained expression and summing up similar terms,
we obtain the equality required in Lemma~2:
\begin{equation}
S_{k_1,...,k_n}A'I_{\bar 0}[\psi]=\frac{k_B T}{\hbar}\biggl(
-i\frac{\partial^2 V'[x']}{\partial x^{'2}_j}\psi[x']+i\psi[x']
\biggr).
\end{equation}

Now we are ready to compute
$S_{\bar 0}A' B^{'-1}A'I_{\bar 0}$ using Lemmas 1 and~2, by formula~(\ref{S_0AB}).
By Lemma~1 the summands in formula~(\ref{S_0AB}) can be nonzero only if $k_j=2$ for some $j\in\{1,...,n\}$,
and the rest of $k_1,...,k_n$ equal~0.  Lemmas 1 and~2 yield expressions for operators $S_{\bar 0}A'I_{k_1, ...,k_n}$
and $S_{k_1, ...,k_n} A'I_{\bar 0}$ in this case. So, by formula (\ref{S_0AB}) and Lemmas 1,~2 we have:
\begin{eqnarray} \label{S_0AB2}
S_{\bar 0}A' B^{'-1}A'I_{\bar 0}[\psi]
&=&
-\!\!\!\!\!\!\!\!\!\!\!\sum_{\stackrel {k_1, ...,k_n=0}{k_1+ ...+k_n=k\geq1}}^\infty\!\!\!\!\!\!\!\!\!\!\!
k^{-1} S_{\bar 0}A'I_{k_1, ...,k_n}S_{k_1, ...,k_n} A'I_{\bar 0}[\psi]\nonumber\\
&=&
-\!\!\!\!\!\!\!\!\!\sum_{\stackrel {k_1, ...,k_n\in\{0; 2\}}
{k_1+ ...+k_n=2}}\!\!\!\!\!\!\!\! 2^{-1} S_{\bar 0}A'I_{k_1, ...,k_n}S_{k_1, ...,k_n} A'I_{\bar 0}[\psi]\nonumber\\
&=&
-\sum_{j=1}^n 2^{-1}\frac{i}{2}\frac{k_B T}{\hbar}
\frac{k_B T}{\hbar}\biggl(
-i\frac{\partial^2 V'}{\partial x^{'2}_j}\psi+i\psi
\biggr)
\nonumber\\
&=&
-\frac{1}{4}\biggl(\frac{k_B T}{\hbar}\biggr)^2
\biggl(\sum_{j=1}^n\frac{\partial^2 V'}{\partial x^{'2}_j}\psi-n\psi
\biggr).
\end{eqnarray}

If in this expression we return to the initial coordinates by formula (\ref{new_var}), then we obtain the equality
\begin{equation}
S_{\bar 0}A' B^{'-1}A'I_{\bar 0})[\psi]=
-\frac{1}{4}\biggl(\frac{1}{m}
\sum_{j=1}^n\frac{\partial^2 V}{\partial x^{2}_j}-
\biggl(\frac{k_B T}{\hbar}\biggr)^2
n\biggr)\psi.
\end{equation}
Thus, we have computed the operator in the second summand in equation~(\ref{eq_diff'_1_psi}).
This operator gives, after multiplying by $-ih\gamma^{-1}$, the second summand in the operator
$\hat H_1$  from Theorem~4.

{\bf Acknowledgments.} The author is thankful to Professor G.~L.~Litvinov for support of the work in this direction,
and mourns for loss of this wonderful person and friend.

\end{document}